\documentclass[manuscript]{emulateapj}

\newcommand{\msun}{M$_{\odot}$}

\shorttitle{Fluorine variations in M22}
\shortauthors{D'ORAZI ET AL.}

\begin{document}


\title{Fluorine variations in the globular cluster NGC~6656 (M22): \\
implications for internal enrichment timescales\footnote{Based on observations taken with ESO telescopes under program 087.0319(A)}}


\author{Valentina D'Orazi\altaffilmark{1,2}}
\author{Sara Lucatello\altaffilmark{3}} 
\author{Maria Lugaro\altaffilmark{2}} 
\author{Raffaele G. Gratton\altaffilmark{3}} 
\author{George Angelou\altaffilmark{2}}
\author{Angela Bragaglia\altaffilmark{4}}
\author{Eugenio Carretta\altaffilmark{4}}
\author{Alan Alves-Brito\altaffilmark{5}} 
\author{Inese I. Ivans\altaffilmark{6}}
\author{Thomas Masseron\altaffilmark{7}} 
\author{Alessio Mucciarelli\altaffilmark{8}} 
\altaffiltext{1}{Department of Physics and Astronomy, Macquarie University, Balaclava Rd, North Ryde, NSW 2109, Australia}
\altaffiltext{2}{Monash Centre for Astrophysics (MoCA), School of Mathematical Sciences, Building 28, Monash University, Clayton, VIC 3800, Australia}
\altaffiltext{3}{INAF Osservatorio Astronomico di Padova, vicolo dell'Osservatorio 5, I-35122, Padova, Italy}
\altaffiltext{4}{INAF Osservatorio Astronomico di Bologna, via Ranzani 1, I-40127, Bologna, Italy} 
\altaffiltext{5}{Research School of Astronomy and Astrophysics, The Australian National University, Cotter Road, Weston, ACT 2611, Australia}
\altaffiltext{6}{Department of Physics and Astronomy, The University of Utah, 115 South 1400 East, Salt Lake City, UT 84112-0830, USA}
\altaffiltext{7}{Institut d'Astronomie et d'Astrophysique, Universit\'{e} libre de Bruxelles, Boulevard du Triomphe, 
B-1050, Brussels, Belgium}
\altaffiltext{8}{Dipartimento di Astronomia, Universit\`{a} di Bologna, via Ranzani 1, I-40127, Bologna, Italy}

\email{valentina.dorazi@mq.edu.au}



\begin{abstract}
Observed chemical (anti)correlations in proton-capture elements among 
globular cluster stars are presently 
recognized as the signature of self-enrichment from now extinct, previous generations of stars. 
This defines the multiple population scenario. Since fluorine is also affected by proton captures, determining its abundance in 
globular clusters provides new and complementary clues regarding  the nature of these previous generations, and supplies strong observational constraints to the chemical enrichment timescales.
In this paper we present our results on near-infrared CRIRES spectroscopic observations of six cool giant stars in NGC 6656 (M22): 
the main objective is to derive the F content and its internal variation in this peculiar cluster, which exhibits significant changes in both light- and heavy- element abundances.
We detected F variations across our sample beyond the measurement uncertainties and found that the F abundances are positively correlated with O and anticorrelated with Na, as expected according to the multiple population framework.
Furthermore, our observations reveal an increase in the F content between the two different sub-groups, 
$s$-process rich and $s$-process poor, hosted within M22. The comparison with theoretical models suggests that asymptotic giant stars with masses between 
4 and 5\msun~are responsible for the observed chemical pattern, confirming evidence from previous works: 
the difference in age between the two sub-components in M22 must be not larger than a few hundreds Myr. 

\end{abstract}


\keywords{globular clusters: individual (NGC 6656)  
---stars:abundances ---:stars: AGB and post-AGB--- stars: Population II ---}



\section{Introduction}

Although Galactic globular clusters (GCs) display a distribution in their global 
parameters (e.g., mass, metallicity, concentration, horizontal-branch 
morphology), the internal variation of elements affected by proton captures (hereafter $p$-capture elements) appears a 
ubiquitous feature (\citealt{carretta09a},b). It is clear that GC stars 
exhibit large changes in the C, N, O, Na, Mg, and Al abundances, whereas (in 
archetypical systems at least) internal spreads in iron-peak, heavy 
$\alpha$- (Ca, Ti) and $slow$ neutron-capture ($s$-process) 
elements all remain within 
observational uncertainties (\citealt{carretta09c}, 2010a; 
\citealt{james04}; \citealt{smith08}; \citealt{dorazi10}). The changes in the 
$p$-capture elements give rise to a clear chemical pattern: depletion 
in C, O, and Mg abundances always correspond to enhancements in N, Na, 
and Al (the so-called light-element anticorrelations). This behavior 
bears the evidence of H burning at high temperature and points to the presence of 
multiple stellar generations. It is argued that the ejecta from a 
fraction of first generation of stars (initially C-O-Mg rich, sharing the same
chemical composition of field stars at the same metallicity) mix 
with primordial gas, providing a medium from which the second generation 
stars (C-O-Mg poor and N-Na-Al rich) formed. We refer the reader to 
\cite{gratton12} for a recent, extensive review on this topic. In 
this scenario, the H-burning abundance patterns from the first 
generation stars are imprinted in the second generation and are present 
from birth. The nature of the stars that enriched the intercluster gas remains 
uncertain but possible candidates include intermediate-mass asymptotic 
giant branch (AGB) stars 
undergoing hot bottom burning (HBB, e.g., \citealt{dantona83}; \citealt{ventura01}), fast rotating 
massive stars (e.g., \citealt{decressin07}), massive binaries 
(\citealt{demink09}), and novae (\citealt{smith96}; \citealt{maccarone12}).

Interestingly, this already complex picture is further obfuscated by the 
presence of some peculiar clusters, such as $\omega$ Centauri 
(\citealt{jp10}; \citealt{marino11a}), NGC~1851 (\citealt{yong08}; 
\citealt{carretta10a}), Terzan~5 (\citealt{ferraro09}), NGC~6715 (M54, 
\citealt{carretta10b}), and NGC~2419 (\citealt{cohen11}, see however 
\citealt{mucciarelli12} for a different view). In these GCs, along with 
changes in $p$-capture elements, internal variations in the heavy 
element abundances have been detected. Species ranging from the 
iron-peak (e.g., Fe) to the $s$-process elements (Ba, La) vary 
stochastically from cluster to cluster beyond what is  
expected from observational errors.

The metal-poor GC NGC 6656 (M22, [Fe/H]\footnote{We use the notation [X/H]=A(X)$-$A(X)$_{\odot}$, 
where A(X)=log($\frac{N_{X}}{N_H}$)+12}=$-$1.70 \citealt{harris96} 
-updated in 2010) belongs to this class of GCs, 
and due to its peculiar nature has 
received extensive attention (\citealt{pilachowski82}; 
\citealt{norris83}; \citealt{bw92}; \citealt{kayser08}; 
\citealt{marino09}; \citealt{dacosta09}; \citealt{dacosta10}; \citealt{abrito12}). Recently, 
Marino et al. (2011b,\nocite{marino11b} hereafter MSK11) presented results from their high-resolution 
spectroscopic study of 35 giant stars, deriving abundances for 
iron-peak, $\alpha$, $p$-capture, and neutron-capture elements. 
This detailed abundance analysis provided a unique 
opportunity to investigate the chemical enrichment history of M22.

This GC is comprised of two distinct groups of stars, characterised by an 
offset in metallicity and in $s$-process element content. The first 
group displays a metallicity of $<$[Fe/H]$>$=$-$1.82$\pm$0.02 with 
$<$[$s$/Fe]$>$=$-$0.01$\pm$0.01 and the second group has  
$<$[Fe/H]$>$=$-$1.67$\pm$0.01 with $<$[$s$/Fe]$>$=+0.35$\pm$0.02. Note that 
the [$s$/Fe] ratios were computed by averaging the abundances of Y, Zr, 
Ba, La, Nd (see MSK11 for details). Each of these two sub-groups 
exhibits the classical Na-O and C-N anticorrelations
shown by the archetypical GCs. Given that 
the [Eu/Fe] ratio serves as a $rapid$ neutron-capture ($r$-process) 
tracer\footnote{In the 
Solar System, $\approx$ 97\% of Eu was synthesized via the $r$-process (\citealt{burris00}, and references therein).} and that 
enhancements in the $s$-process elements are not accompanied by a 
similar trend in the [Eu/Fe] ratio, it can be inferred that the 
dichotomy in the $n$-capture abundances may be due to a first generation 
of polluters that produced the $s$-process only. One possible scenario, 
which was advocated by MSK11, is that the $weak$ 
$s$-process component activated in stars with masses larger than 
$\sim$25 M$_{\odot}$ during core He-burning and C-shell phases (\citealt{raiteri93}; \citealt{pignatari10}), may have contributed to the observed 
abundance patterns. However, in a complimentary study, \citet{roederer11} 
focused on the heavy element content (from Y to Th) of six stars across 
the two stellar sub-groups, and ruled out the massive star origin. They 
concluded that the gas from which the second stellar population formed 
was enriched in $s$-process material from a class of relatively massive 
AGB stars  
(M $\approx$ 5 M$_{\odot}$). In these stars, the production of $s$-process 
elements is due to the activation of the 
$^{22}$Ne($\alpha$,n)$^{25}$Mg neutron source, whereas in their 
lower-mass counterparts the main neutron source 
is the $^{13}$C($\alpha$,n)$^{16}$O reaction 
(\citealt{busso99}).

Neither of the proposed scenarios provides a comprehensive explanation 
for all the observed chemical features and we are left with numerous 
unsolved issues. For example, \cite{roederer11} question why the 
$s$-process dichotomy is only present in M22: if massive AGB 
stars are the cause of the GC Na-O anticorrelation, then all clusters 
should present the $s$-process elements correlated with Na and anti-correlated
with O, which is not observed (e.g., \citealt{dorazi10}).

In this paper we turn to an alternative diagnostic. We present fluorine 
abundances for a sample of six cool giant stars in M22, carefully 
selected from both sub-stellar groups as defined by MSK11. 
Fluorine abundances are a powerful tracer of the polluter mass range 
in M22 because the F production is highly dependent 
on the stellar mass.

Theoretical models of AGB stars (e.g.,  \citealt{jorissen92}) predict that F is produced due to the 
activation of the chain of reactions 
$^{18}$O(p,$\alpha$)$^{15}$N($\alpha$,$\gamma$)$^{19}$F in the He intershell 
during the recurrent thermal pulses associated with He burning. During 
the early phases of each thermal pulse, the H-burning ashes are ingested 
in the convective region developing in the He intershell. These ashes 
are rich in $^{13}$C and $^{14}$N and their ingestion in the He-burning 
layer results in production of $^{18}$O due to $\alpha$ captures on 
$^{14}$N. At the same time protons are released by the 
$^{14}$N(n,p)$^{14}$C reaction, with neutrons coming from 
$^{13}$C($\alpha$,n)$^{16}$O. After the quenching of 
each thermal pulse the envelope may sink in mass deep in the He 
intershell and carry $^{19}$F to the convective envelope via a process 
known as the ``third dredge-up'' (TDU). The peak of F production in AGB 
stars is reached for stars of initial masses $\sim$ 2 \msun~\citep{lugaro04}. If the mass of the stars is higher than roughly 
5 \msun, and depending on the metallicity, fluorine is 
destroyed both via $\alpha$ captures in the He intershell, and via 
proton captures at the base of the convective envelope due to HBB. AGB 
stars that experience HBB can also destroy O and Mg and produce Na and 
Al; as a consequence, according to the multiple population scenario, we 
should expect the abundances of F to be correlated with O (and Mg) and 
anti-correlated with those of Na (and Al). This prediction was 
observationally confirmed by \cite{smith05} in the 
intermediate-metallicity GC M4 and by \cite{yong+08} in NGC 6712.

Other sites have also been suggested for F production: the $\nu$-process 
in core-collapse supernovae (\citealt{woosley90}), and core He burning 
in Wolf-Rayet stars (\citealt{meynet00}; \citealt{palacios05}). However, 
low-mass AGB stars are the only site observationally confirmed 
(\citealt{jorissen92}; \citealt{abia10}).

This paper is organized as follows: observations are described in 
Section~\ref{sec:obs}, while details on abundance analyses are given in 
Section~\ref{sec:analysis}. Our results are then presented in 
Section~\ref{sec:results} and discussed in Section~\ref{sec:discussion}. 
A summary closes the manuscript (Section \ref{sec:summary}).

\section{Observations}\label{sec:obs}

Our sample includes six giant stars, for which stellar parameters (T$_{\rm eff}$, log$g$, [Fe/H] and microturbulence $\xi$) 
along with $p$-capture and $s$-process element abundances were derived by MSK11. 
We selected three stars belonging to the metal-poor component (MP, also $s$-process poor) and three stars 
from the metal-rich (MR, $s$-process rich) one. Within each of these sub-groups we also selected both O-rich (Na-poor) and 
O-poor (Na-rich) stars, spanning a range from [O/Fe]=+0.11 to [O/Fe]=+0.48 dex. 

The main objective of our investigation was the determination of the fluorine abundances. 
Observationally, F (whose only stable isotope is $^{19}$F) is difficult to detect spectroscopically; 
the only atomic lines (the ground-state transitions of F~{\sc i}) 
that might be revealed lie in the far UV. On the other hand, HF molecular transitions 
are easily observable in the near-infrared (around $\sim$23000 \AA). 
Our analysis focuses on the HF(1$-$0) R9 transition, located at $\lambda$23358.3 \AA. Although not being the strongest 
feature in this wavelength range, this line
is considered to be one of the best abundance indicator for F, because it is free of blends (\citealt{abia09}; \citealt{lucatello11}). 

\begin{table*}
\caption{Information on Target Stars}\label{t:log}
\begin{center}
\begin{tabular}{lcccccr}
\hline\hline
Star   &     RA      &      Dec       &   V    &  K   & Exposures & S/N \\
       &  (hh:mm:ss) &   ($^{o}$:':'')&  (mag) &  (mag) & (s)  &    \\
       &             &                &        &      &          &           				\\
\hline   
       &             &                &        &       &           &        					\\
IV-97  & 18:36:41.06 &	 $-$23:58:18.9 & 11.043 & 6.759 &  10x90s  &	300 \\
III-14 & 18:36:15.10 &	 $-$23:54:54.6 & 11.134 & 6.743 &  17x120s  &   400  \\ 
III-15 & 18:36:15.61 &	 $-$23:55:01.2 & 11.362 & 7.138 &   8x180s  &   300 \\
C      & 18:36:10.21 &	 $-$23:48:44.0 & 11.309 & 6.737 &   6x180s  &   300  \\
III-52 & 18:36:10.18 &	 $-$23:54:21.8 & 11.526 & 7.459 &  10x180s  &   350 \\
V-2    & 18:36:28.02 &	 $-$23:55:01.6 & 11.498 & 7.276 &  14x120s  &   350 \\
       &             &               &        &       & 	  &    \\
\hline\hline
\end{tabular}
\end{center}
\end{table*}

High-resolution, near-infrared spectroscopic observations were carried out in service mode with CRIRES (CRyogenic high-resolution 
InfraRed Echelle Spectrograph, \citealt{kaeufl04}) located at VLT UT1 on 2011 April, July and August 
(program: 087.0319(A), PI: VD). In Table~\ref{t:log}, we list information on target stars, reporting identifications, magnitudes
(see MSK11 for details), exposure times, and S/N ratios per pixel around the HF feature. 
Note that our selection was limited to the cooler stars in the MSK11's sample due to our imposed requirement of 
relatively strong HF lines. 
We also observed several (hot) early-type stars before and/or after of each target observations, 
in order to remove telluric contamination from our scientific frames.

We employed the 0.4$^{''}$ slit and the grating order \#24, achieving a resolution of $R$$\sim$50,000
and a wavelength coverage from $\lambda$22948.5 \AA~ to $\lambda$23410.3 \AA.  
This allowed us to include the HF(1$-$0) R9 line, numerous $^{12}$C$^{16}$O vibration/rotation lines (used to derive C abundances) 
and the Na~{\sc i} line at $\lambda$23379 \AA.

\begin{center}
\begin{figure}
\includegraphics[width=8cm]{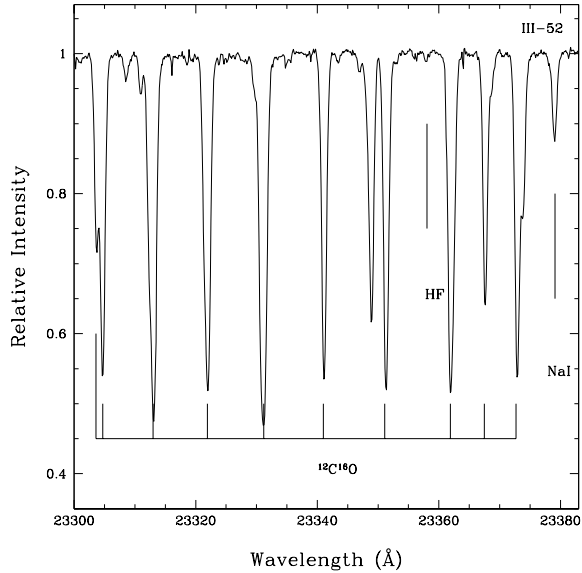}
\caption{Example of a spectrum for star III-52 }\label{f:spectra}
\end{figure}
\end{center}

Data reduction was accomplished by means of the CRIRES pipeline (version 2.4), running under the {\sc gasgano} 
environment\footnote{http://www.eso.org/sci/software/gasgano/}, which provides one-dimensional, wavelength calibrated spectra. 
Telluric feature subtraction, rest-frame translation and continuum normalization were then carried out within 
IRAF\footnote{IRAF is the Image Reduction and Analysis Facility, a general purpose software system
for the reduction and analysis of astronomical data. IRAF is written and supported by National Optical Astronomy 
Observatories (NOAO) in Tucson, Arizona.}. 
An example of our spectra is shown in Figure~\ref{f:spectra} for star III-52; HF, CO and Na~{\sc i} lines are marked.

\section{Abundance analysis}\label{sec:analysis}

Fluorine abundances were determined through spectral synthesis using the MOOG code (\citealt{sneden73}, 2011 version) and the 
Kurucz (1993\nocite{kur93}) set of stellar atmosphere models (with no overshooting) as in MSK11's analysis, from which we
retrieved stellar parameters as well as O and $s$-process element abundances. 
However, had we instead adopted the MARCS grid (\citealt{gustafsson08}), the difference on the resulting abundances 
would have been less than 0.04 dex.

Concerning the HF feature, we took as excitation potential (EP) $\chi$=0.227 eV (\citealt{decin00}) 
and a log$gf$=$-$3.971 (\citealt{lucatello11}). This last value is very close to that used in previous studies 
focusing on F abundance determination in GCs (e.g., \citealt{smith05}; \citealt{yong+08}; \citealt{abrito12}), 
log$gf$=$-$3.955 (from \citealt{jorissen92}). 
On the other hand, our EP is about 0.25 eV lower compared to
those works: this difference implies an offset between our abundances and those previously published in the literature
of roughly 0.30 dex (see also the discussion in Section~\ref{sec:alan}).
Furthermore, we derived C abundances. For this purpose, we assumed O values from the optical range given by MSK11, 
since our spectra did not cover any suitable OH line, whose stronger transitions extend in the $H$ band (around $\sim$15000\AA). 
The CO line lists come from B. Plez (private communication). 
Finally, for the Na~{\sc i} line at $\lambda$23379, atomic parameters  ($\chi$=3.750 eV; log$gf$=0.530) were 
taken from VALD\footnote{Vienna Atomic Line Database 
(www.astro.uu.se/~vald/php/vald.php)}.

As a first step, we checked our line list on the infra-red atlas of the Arcturus spectrum (\citealt{hinkle}, 
available at ftp://ftp.noao.edu/catalogs/arcturusatlas/). 
Assuming a T$_{\rm eff}$=4286 K, log$g$=1.67, $\xi$=1.74 kms$^{-1}$ and 
[Fe/H]=$-$0.52 (following \citealt{ramirez11}) we obtained 
an [F/Fe]=$-$0.15\footnote{We used as solar fluorine abundances A(F)$_{\odot}$=4.56 (\citealt{ag89}; \citealt{asplund05}).}, 
to be compared to the value given by \cite{abia09} of [F/Fe]=0.10 dex. 
The difference is completely explained by the higher EP adopted in that study.

Moreover, we inferred a C abundance of A(C)=8.01 
(under the assumption that A(O)=8.81, that is [O/Fe]=0.40, with a solar abundance of A(O)$_{\odot}$=8.93), which is in very good 
agreement with values of \cite{abia09} (i.e., A(C)=8.06) and \cite{ryde10} (A(C)=8.08). 
As for C solar abundance we adopted the value of A(C)$_{\odot}$=8.56, leading to [C/Fe]=$-$0.03.

Finally, from the Na~{\sc i} feature at 23379\AA, we derived A(Na)=6.01 which results in [Na/Fe]=+0.2 dex (setting A(Na)$_{\odot}$=6.33).

\begin{center}
\begin{figure}
\includegraphics[width=8cm]{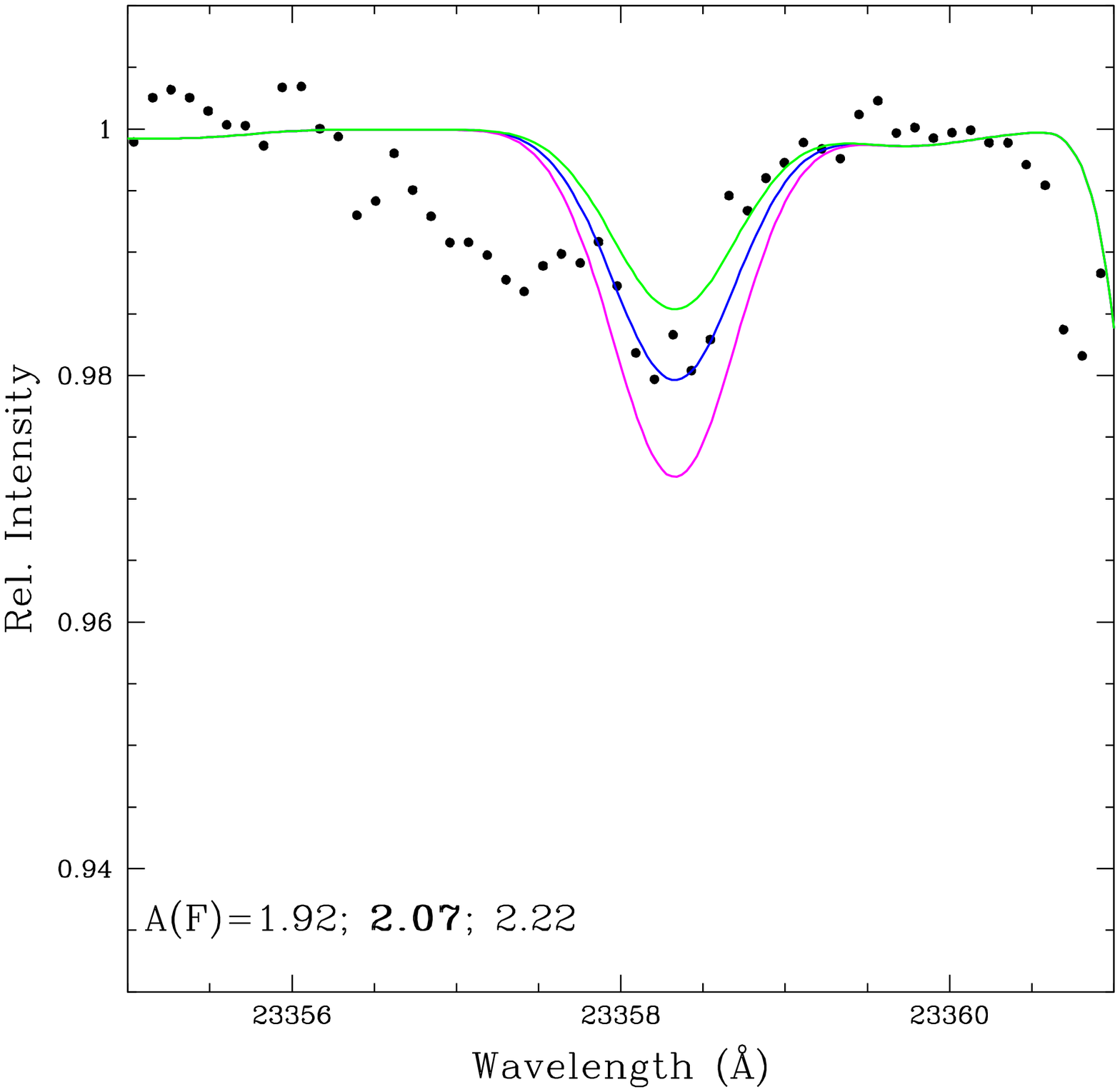}
\caption{Synthesis of the HF feature for star C.}\label{f:synth1}
\end{figure}
\end{center}

\begin{center}
\begin{figure}
\includegraphics[width=8cm]{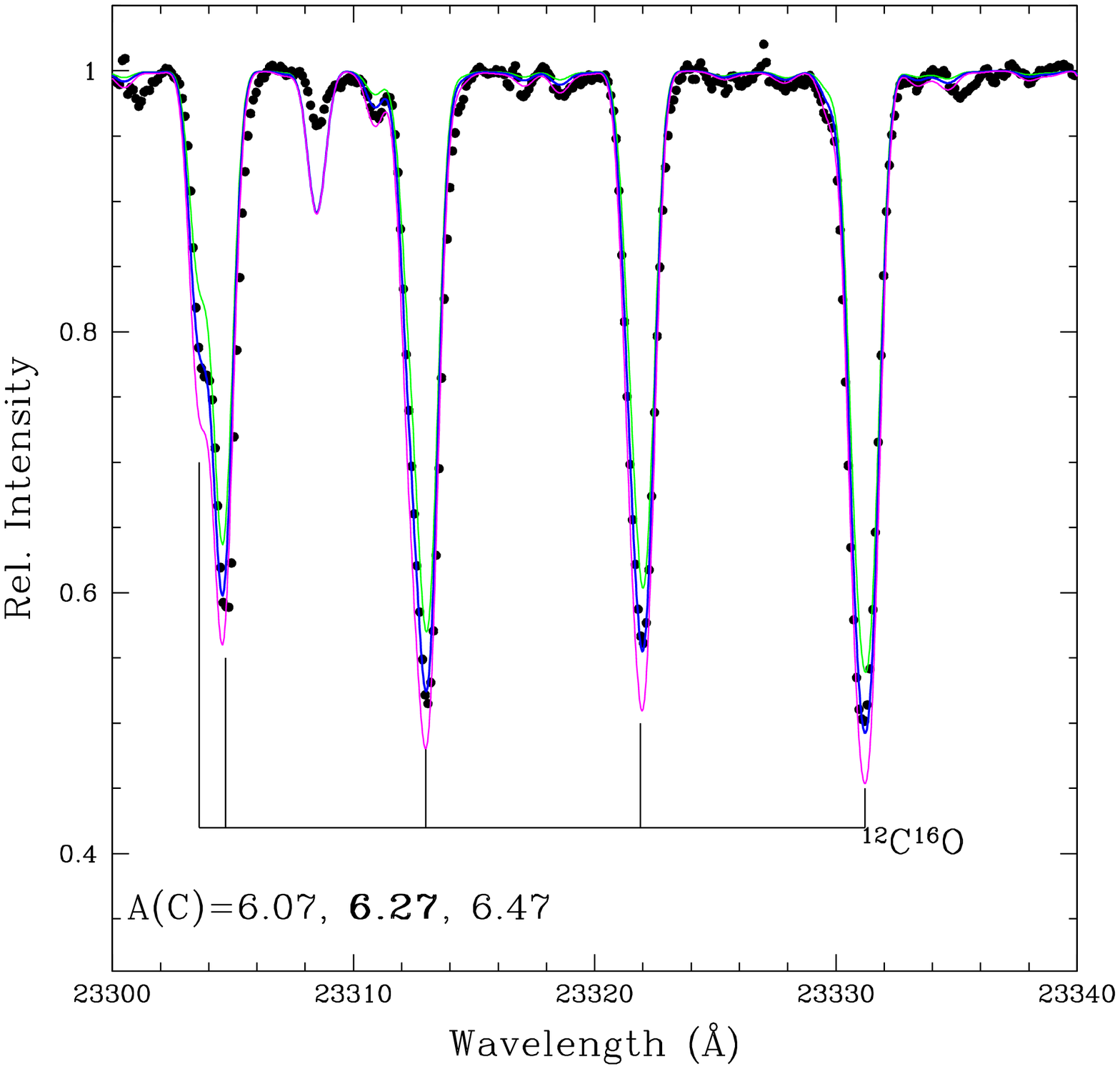}\\
\includegraphics[width=8cm]{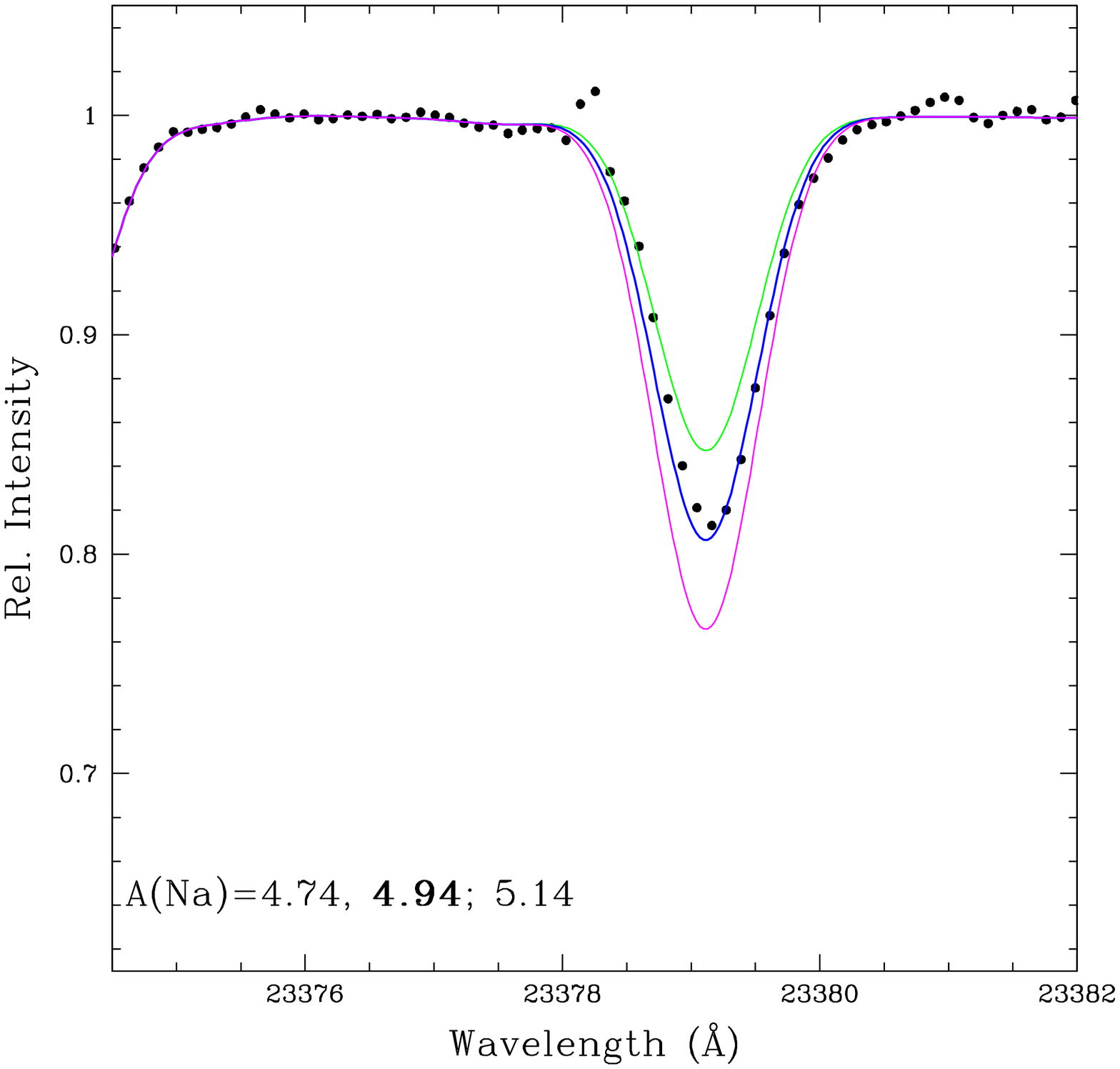}
\caption{Same as Figure~\ref{f:synth1} but for CO (upper panel) and Na (lower panel)} \label{f:synth2}
\end{figure}
\end{center}

Comparison between synthetic and observed spectra were carried out in 
a similar way for our sample stars; an example of spectral synthesis is given in 
Figures~\ref{f:synth1} and~\ref{f:synth2} for star C.

The sensitivity of the F abundance to input stellar parameters 
was evaluated by separately changing effective temperature, surface gravity and 
microturbulence values.
The intensity of the synthetic line of the HF is
particularly sensitive to the adopted T$_{\rm eff}$, the other parameters affecting it at a lower degree
(see also \citealt{abia09}; \nocite{abia11}2011).
A change of $\Delta$(T$_{\rm eff}$)=+70 K, $\Delta$(log$g$)=+0.15, and $\Delta$($\xi$)=+0.13 km s$^{-1}$
(conforming to error estimates given in Table 4 of MSK11) results in a difference in A(F) of 
+0.10, 0.02 and $-$0.02 dex, respectively. The variation of the input metallicity in the model atmosphere has instead a negligible effect.
These are the typical uncertainties that we then summed in quadrature providing a total error, due to stellar parameters, 
of 0.11 dex in our [F/H] ratios. 
Errors due to the best-fit determination (related to the S/N of the spectra and including uncertainties due to the continuum placement) are instead $\pm$0.07 dex. However, we caution the reader that this value should be treated as a lower limit, since the impact of the telluric correction on this considerably weak feature is significant 
(see Section~\ref{sec:results}).  

The total internal error in [F/H] is then obtained adding in quadrature both uncertainties, resulting in 0.13 dex. 

Finally, as far as C and Na are concerned, the typical uncertainties are 0.12 and 0.10 dex, respectively.

\section{Results}\label{sec:results}

\begin{table*}
\caption{Stellar Parameters and Elemental Abundances}\label{t:results}
\begin{center}
\begin{tabular}{lccccccccr}
\hline\hline
Star  &  T$_{\rm eff}$$^{*}$ & $\log g$$^{*}$   & [Fe/H]$^{*}$  & $\xi$$^{*}$        & [O/Fe]$^{*}$  
& $s$-rich$^{*}$   & [C/Fe] & [F/Fe] & [Na/Fe] \\  
      &    (K)         &  (cms$^{-2}$) &        & (kms$^{-1}$) &      &              &        &        &           \\
\hline         
      &                &                     &        &             &        &            &        &        &            \\
IV-97  &  4000    &	    0.05 & $-$1.94   & 2.00  & 0.40  & no  & $-$1.10  &	$-$0.70 & $-$0.40	 \\
III-14 &  4030    &	    0.35 & $-$1.82   & 2.15  & 0.48  & no  & $-$1.10  &	$-$0.80 & $-$0.30	 \\
III-15 &  4070    &	    0.40 & $-$1.82   & 1.85  & 0.11  & no  & $-$1.15  &	$-$1.00 &  0.34	 \\
       &          &              &           &       &       &     &      &          &               \\   
C      &  3960    &	    0.30 & $-$1.69   & 2.25  & 0.25  & yes & $-$0.60  &	$-$0.80 &  0.30	 \\
III-52 &  4075    &	    0.60 & $-$1.63   & 1.75  & 0.45  & yes & $-$0.10  &	$-$0.60 &  0.05	 \\
V-2    &  4130    &	    0.65 & $-$1.57   & 1.75  & 0.15  & yes & $-$0.40  &	$-$0.90 &  0.31	 \\
       & 	  &	         &	     &	     &	     &	   &	      &	      &	\\
\hline\hline
\multicolumn{1}{l}{$^{*}$ from MSK11.}
\end{tabular}
\end{center}
\end{table*}

Our results are shown in Table~\ref{t:results}, where we report stellar parameters and abundances from MSK11 along
with our estimates for F, C, and Na.
Even within our quite limited sample (six stars), we found that 
{\em the F abundance shows a large star-to-star variation}, ranging from [F/H]=$-$2.82 dex to [F/H]=$-$2.23 dex
(i.e., a factor of $\sim$4). The average abundance is $<$[F/H]$>$=$-$2.55$\pm$0.08 (rms=0.20), implying that
the amplitude of this change is beyond the measurement uncertainties (see Section~\ref{sec:analysis}). Moreover, taking into account 
the typical errors for O and F abundances, our study suggests that
the F variation is comparable with that of O, as also found by \cite{yong+08} in the GC NGC~6712.
\begin{center}
\begin{figure}
\includegraphics[width=8cm]{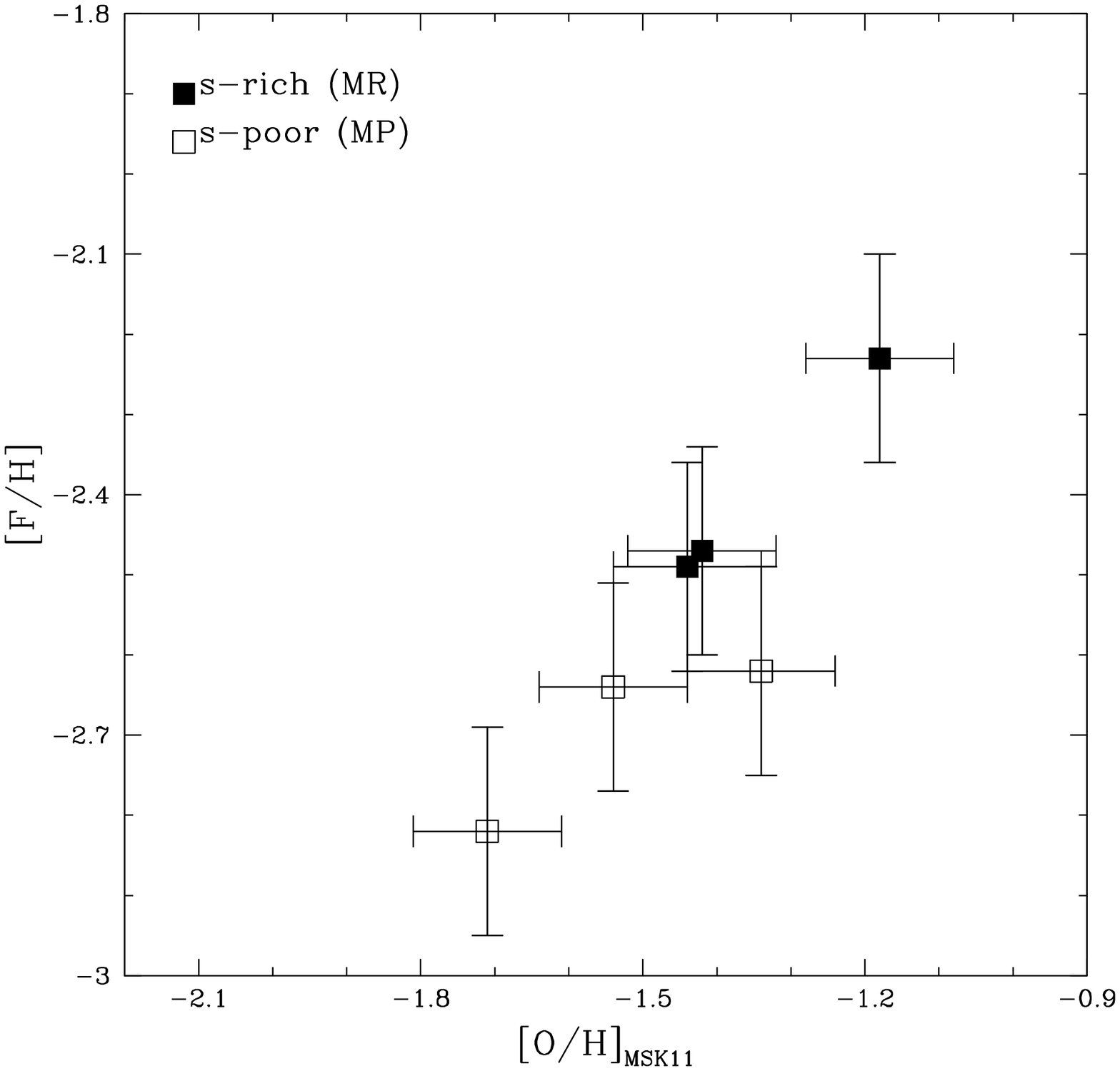}
\caption{Fluorine abundances ([F/H]) as a function of [O/H].}\label{f:OF}
\end{figure}
\end{center}

\begin{center}
\begin{figure}
\includegraphics[width=8cm]{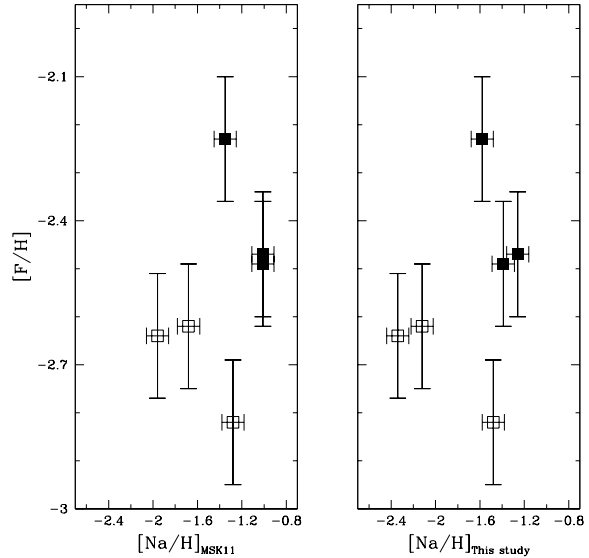}
\caption{[F/H] versus [Na/H] from MSK11 (left panel) and from this study (right panel). 
Symbols are as in Figure~\ref{f:OF}.}\label{f:NaF}
\end{figure}
\end{center}

In Figures~\ref{f:OF} and~\ref{f:NaF} we plot [F/H] ratios as a function of [O/H] and [Na/H], respectively. 
As one can see, F abundances are positively correlated with O: considering the whole sample, 
the Pearson correlation coefficient results in 
$r$=0.89, with a probability to be random smaller than 2\%. 
Focusing on the F-Na diagram (Figure~\ref{f:NaF}), there is the hint for a F-Na anticorrelation, but 
in our small sample the linear correlation 
coefficient is not statistically meaningful. 
However, this does not prove the lack of an anticorrelation, because we are dealing 
with small numbers (only six points), heavily reducing the power of statistical tests. 
In addition, there is no a priori reason why we should combine the two sub-groups as far as the F-Na plane is concerned, 
because we do not expect that they must behave in the same way. If we look at each component separately the presence of a 
F-Na anticorrelation is indeed much more evident (given that we have only three points for each group, it is not meaningful to 
perform a statistical test separately).

The observed chemical pattern can be explained as the evidence of H-burning at high temperature, via the CNO cycle, 
which causes the destruction of F, in conjunction with O depletion and Na enhancement.
More interestingly, those (anti)correlations are revealed in each of the M22 sub-components (the $s$-rich and $s$-poor groups); 
the implications of this finding are discussed in detail in Section~\ref{sec:discussion}.

\begin{center}
\begin{figure}
\includegraphics[width=8cm]{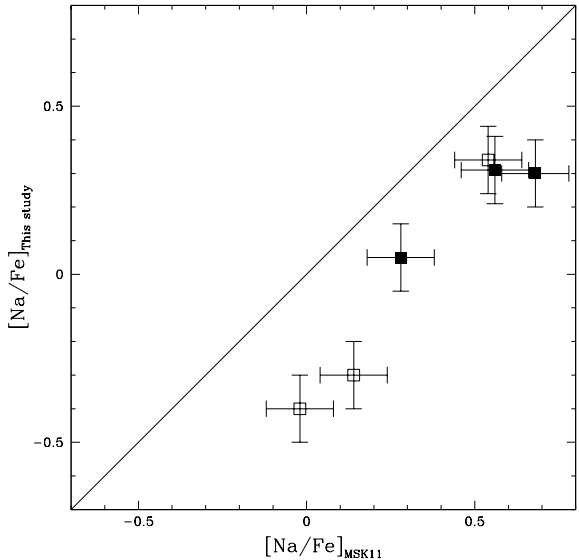}
\caption{[Na/Fe] from the optical range by MSK11 and from this study}\label{f:NaIROpt}
\end{figure}
\end{center}

Regarding Na, we show both our estimate from near-infrared spectroscopy as well as LTE abundances from the optical range by 
MSK11 (right and left-hand panels of Figure~\ref{f:NaF}). 
A difference of $\Delta$([Na/Fe])=0.31$\pm$0.04 (rms=0.09) dex is found between the two 
estimates (see Figure~\ref{f:NaIROpt}) where we compare the two measurements; non-LTE effects can totally account for such a discrepancy (e.g., \citealt{lind11}). 
The total average Na abundance is $<$[Na/Fe]$>$=0.05$\pm$0.13 (rms=0.33); considering separately the two groups we obtain
instead a $<$[Na/Fe]$>$$_{s-\rm poor}$=$-$0.12$\pm$0.23 and 
$<$[Na/Fe]$>$$_{s-\rm rich}$=+0.22$\pm$0.08, which implies $\Delta^{\rm rich}_{\rm poor}$[Na/Fe]=0.34$\pm$0.17 dex. 
This value is in good agreement with that derived by MSK11, based on the whole sample of 35 giants, being 
$\Delta^{\rm rich}_{\rm poor}$[Na/Fe]=0.23$\pm$0.07.

Finally, our sample displays an average C abundance of $<$[C/Fe]$>$=$-$0.74$\pm$0.18, with 
$<$[C/Fe]$>$$_{s-\rm poor}$=$-$1.12$\pm$0.02 and $<$[C/Fe]$>$$_{s-\rm rich}$=$-$0.37$\pm$0.14.
Thus, on average, $s$-rich stars exhibit larger C abundances, which qualitatively agrees with previous works
(e.g., \citealt{brown90}). However, while the difference between the two groups from the optical CH bands derived
by MSK11 is $\Delta^{\rm rich}_{\rm poor}$[C/Fe]=0.35$\pm$0.13 dex, we achieved a 
much larger value of $\Delta^{\rm rich}_{\rm poor}$[C/Fe]=0.75$\pm$0.10 dex. 
The same conclusion was reached by \cite{abrito12} who found a variation of 
$\Delta^{\rm rich}_{\rm poor}$[C/Fe]=0.78$\pm$0.15 dex, 
from high-resolution, near-infrared spectroscopy of nine cool giants (see also Section~\ref{sec:alan}). 
The reason of such a difference in C abundance from the optical and from the near-infrared is not clear and no 
obvious trends with stellar parameters (e.g., temperature, gravity, microturbulence, and/or metallicity) 
seem to be present. 
Further investigations are needed to explore this issue. 
\subsection{Comparison with \cite{abrito12}}\label{sec:alan}

In a recent paper, \cite{abrito12} carried out high-resolution 
($R$=50,000), near-infrared (both $H$ and $K$ bands) spectroscopic 
observations with Phoenix@Gemini-South of nine red giant branch (RGB) stars in M22. They investigated their F 
content, presenting also C, N, O, Na and Fe abundances.

\begin{center}
\begin{figure}
\includegraphics[width=9cm]{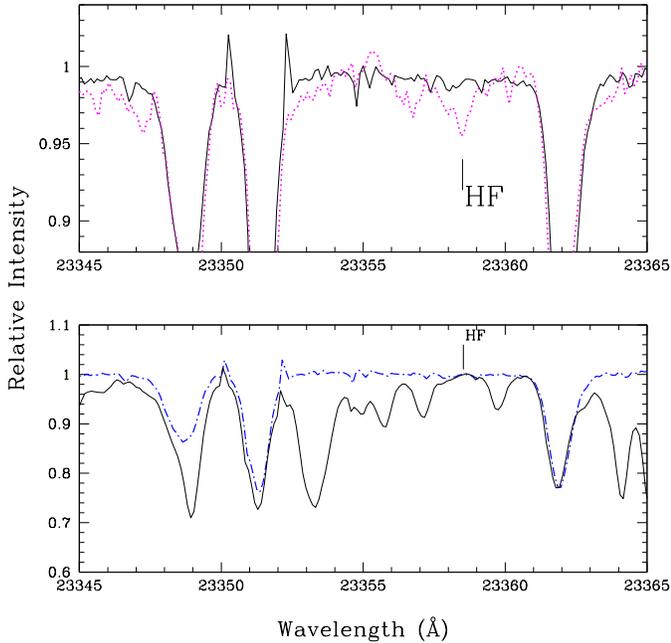}
\caption{Upper panel: comparison of our spectrum for star III-15 (solid line) with the one by \cite{abrito12} (dotted line).  
Lower panel: superimposition of our spectra for the same sample star with and without the telluric line 
subtraction (dot-dashed and solid lines, respectively).}\label{f:comp_spectra}
\end{figure}
\end{center}

\begin{center}
\begin{figure}
\includegraphics[width=9cm]{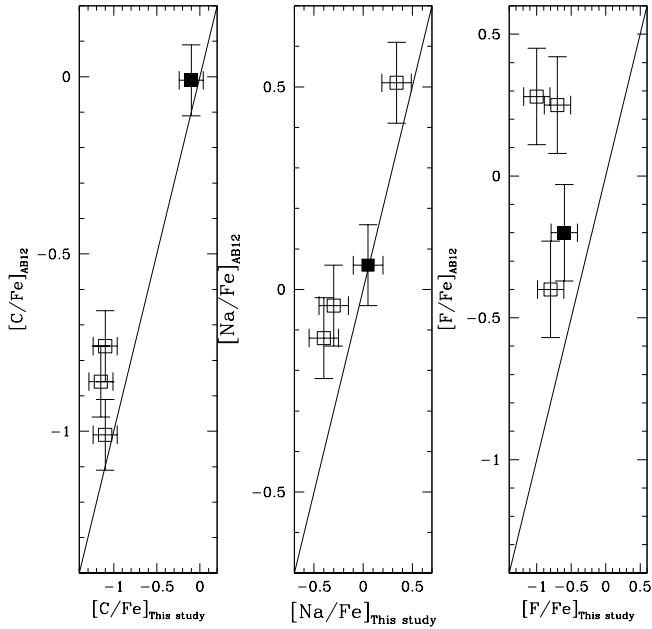}
\caption{Comparison for the four stars in common with Alves-Brito et al. (2012, here labeled as AB12). 
Symbols are as in Figures~\ref{f:OF}, \ref{f:NaF}, \ref{f:NaIROpt}.}\label{f:comp}
\end{figure}
\end{center}

Four of our six stars are in common with that study, namely, 
III-14, III-15, III-52 and IV-97. For stars III-14 and III-52, those 
authors inferred [F/Fe]=$-$0.40 and [F/Fe]=$-$0.20 dex, respectively, 
while we obtained [F/Fe]=$-$0.80 and [F/Fe]=$-$0.60 dex. The adopted EP 
value can easily justify this divergence, accounting for about 0.30 dex 
(see Section~\ref{sec:analysis}); the source of the remaining $\sim$0.1 
dex can be ascribed to the continuum placement, which is critical in 
determining abundances from such a weak line. On the other hand, for 
stars III-15 and IV-97 \cite{abrito12} obtained [F/Fe]=0.28 and 
[F/Fe]=0.25 dex, entailing discrepancies with our estimates 
larger than a factor 10. Note that stellar parameters (T$_{\rm eff}$, log$g$, and $\xi$) are the 
same in both works, as they come from the analysis of MSK11; for the input metallicity, Alves-Brito et al.
used instead their own values coming from the IR spectroscopy and showing an offset of +0.13 dex
compared to the optical ones.  
We investigated the nature of this substantial discordance, and 
attributed it to the telluric feature subtraction. In the upper panel of 
Figure~\ref{f:comp_spectra}, we directly compare our spectrum for 
star III-15 (solid line) with that used by Alves-Brito et al. (dotted 
line\nocite{abrito12}). Their spectrum presents 
stronger features in the vicinity of HF line, 
features expected from telluric contribution. To completely remove the contamination, Alves-Brito et al. 
realized that they needed early-type star targets before and/or after each scientific frame but the logistics 
of their run made this very difficult. 
We instead could observe such targets: the correction to our data, in turn affects the placing of the continuum and 
removes many strong features (as shown in the lower panel of Figure~\ref{f:comp_spectra}). 
As expected, such an effect is significant for the
HF feature, due to its intrinsic weakness, but only marginally affects the C and Na abundance determinations (due to
the strength of their lines), and hence most of the conclusions of that paper. 
This is shown in Figure~\ref{f:comp}, where we plot our [X/Fe] ratio 
as a function of those from \cite{abrito12} for the four stars in common: C 
and Na are comparable between the two studies, with differences of $\Delta$[C/Fe]=+0.20$\pm$0.17 dex
and $\Delta$[Na/Fe]=+0.18$\pm$0.11 dex (in the sense Alves-Brito's study minus our values); 
if we take into account the offsets in [Fe/H] and in the adopted solar 
abundances (they assumed A(C)$_{\odot}$=8.42 and A(Na)$_{\odot}$=6.17), 
those values become $\Delta$[C/Fe]=+0.19$\pm$0.17 and $\Delta$[Na/Fe]=+0.15$\pm$0.11 dex. 
On the other hand, discrepancies in F are significant and cannot be recovered from the different EP and/or solar abundances, 
being for the whole sample of $\Delta$[F/Fe]=+0.75$\pm$0.19 dex.

\section{Discussion}\label{sec:discussion}

\begin{center}
\begin{figure}
\includegraphics[width=8cm]{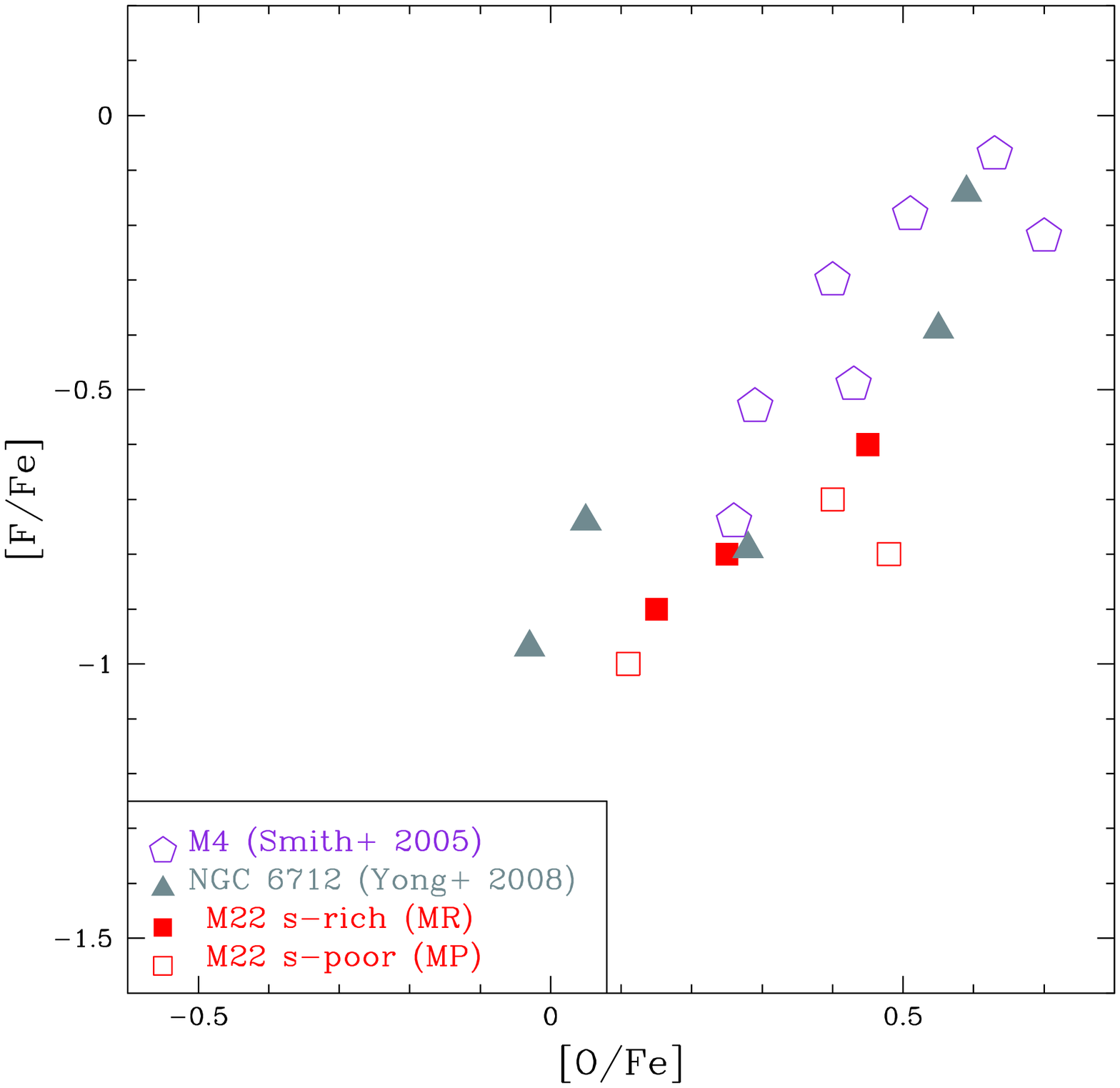}
\caption{[F/Fe] versus [O/Fe] for M22 (this study), M4 (\citealt{smith05}), and NGC 6712 (\citealt{yong+08}).}\label{f:gcs}
\end{figure}
\end{center}

Our main result is the detection of F variations across our sample 
significantly larger than the observational uncertainties. As shown in Figures~\ref{f:OF} and \ref{f:NaF}, the changes in the F 
abundances are correlated with O and anticorrelated with Na. 
This chemical pattern qualitatively matches the 
predictions from the multiple population scenario, according to which 
the stellar ejecta from which  
second generation stars formed carry the signature of hot H burning
causing enhancements in Na (N and Al) and depletions in O and F (C and
Mg). The F-O diagram presented in Figure~\ref{f:gcs} demonstrates that 
M22 shares a similar behaviour as M4 and NGC6712, the other two GCs 
for which F has been explored\footnote{\cite{cunha03} presented F 
abundances for two giants in $\omega$ Cen. However,  
they provide an F measurement only for star ROA219, giving 
an upper limit for star ROA 324. Discussion related to the internal F 
variation in this peculiar 
GC is still not possible with the currently available measurements. For 
this reason we acquired CRIRES spectra of 12 $\omega$~Cen giants; 
results will be presented in a forthcoming paper (S. Lucatello et al., in 
preparation).}.

Furthermore, the F-O-Na (anti) correlations can be marked separately within 
each of the two sub-groups enclosed in M22 as clearly illustrated in 
Figures~\ref{f:OF} 
and~\ref{f:NaF}, where the $s$-process poor and the $s$-process rich stars are 
labeled with empty and filled symbols, respectively. The same 
conclusion was drawn by MSK11 when considering the Na-O and 
C-N planes.

\begin{center}
\begin{figure}
\includegraphics[width=8cm]{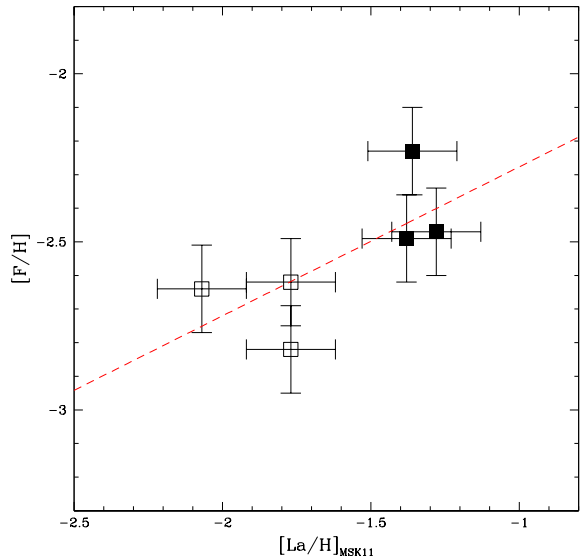}
\caption{[F/H] versus [La/H]. Symbols as in Figures~\ref{f:OF}, \ref{f:NaF}, \ref{f:NaIROpt}, and \ref{f:comp}. The dashed line is a least-squares fit to data points.}\label{f:LaF}
\end{figure}
\end{center}

Very interestingly, beyond the internal spread in F characterizing each 
sub-component, we measured an increase in the F content between the two 
different stellar generations in M22. {\em The $s$-process-rich group 
has, on average, larger F abundances than the $s$-process-poor group.} 
This is shown in Figure~\ref{f:LaF}, where we plot our F abundances 
([F/H]) as a function of [La/H] from MSK11. There is a 
positive correlation between the two ratios, suggesting that  
the polluters responsible for the $s$-process production  
must account for a simultaneous F production. 

There are two classes of objects producing both $s$-process elements
and fluorine. The first one is very massive stars (mass roughly $>$
40\msun). These produce F in the initial phases of core He
burning and expel it in the interstellar medium via winds during the
Wolf-Rayet phase (\citealt{meynet00}). They also produce $s$-process
elements during core He and shell C burning (e.g., \citealt{pignatari10}).
Production of both F and $s$-process elements in these massive stars
depends on the initial CNO abundances and thus decreases with the
stellar metallicity. Inclusion of stellar rotation enhances the
$s$-process production at low metallicity
(\citealt{pignatari08}; \citealt{chiappini11}); however, it appears to
decrease the production of fluorine (\citealt{palacios05}).
One problem already stressed by \cite{roederer11} 
when considering these stars is that there is no reason why the SNe that enriched the $s$-process rich group host the $weak$ 
component and those that polluted the $s$-process poor do not (see Roederer et al. for details\nocite{roederer11}).

The second class of objects producing both F and $s$-process elements 
are AGB stars (\citealt{forestini92}; 
\citealt{jorissen92}; \citealt{mowlavi98}; \citealt{karakas03}; 
\citealt{lugaro04}, 2012\nocite{lugaro12}).
To get deeper insights into the nature of the candidate AGB stars possibly responsible 
for the observed abundance trends in M22 we compare our results with the 
recent set of models by \cite{lugaro12}. They presented AGB models for 
masses 0.9$-$6.0~\msun~and metallicity three times lower than that of 
the cluster under consideration (i.e., [Fe/H]=$-$2.3 dex).
 
From our data we infer that there is an increase of $\Delta$[F/H]$^{s-rich}_{s-poor}$=+0.40$\pm$0.15 dex in 
the fluorine content between the two groups. This 
estimate was done by taking into account the F content of the O-rich 
stars only, because they do not show any depletion due to the HBB. We 
averaged the F abundances in stars IV-97 and III-14, both belonging to the 
$s$-poor group, finding a mean value $<$[F/H]$>$=$-$2.63$\pm$0.01 dex. Since we have 
only one O-rich star in the $s$-rich group, III-52, we chose it as 
representative of the F abundance for the group, that is [F/H]=$-$2.23$\pm$0.15 dex.
The F increase of +0.40 dex is accompanied by a corresponding 
enhancement in La of $\Delta$[La/H]$^{s-rich}_{s-poor}$=0.56 $\pm$0.18 dex, since the values are 
$<$[La/H]$>$=$-$1.92$\pm$0.15 dex and [La/H]=$-$1.36$\pm$0.10 dex, respectively, for the two 
groups.

Comparing these values with the model predictions by \cite{lugaro12} we 
found that AGB stars with masses of $\approx$ 4$-$5 \msun\ can well 
reproduce the observed pattern. Lower-mass AGB models do not fit our 
observational requirements because they over-produce fluorine. This is 
true even if we consider that these AGB model predictions are roughly one 
dex too high to match the observation of carbon-enhanced metal-poor (CEMP) stars by 
\cite{lucatello11}. Production of F in the $\approx$ 4$-$5 \msun\ 
mass range depends on the delicate balance between the operation of the 
TDU and of HBB. These stars suffer HBB and destroy F during the early 
phases of their AGB evolution, however, toward the end of the evolution, as 
the mass of the convective envelope decreases, HBB ceases while the TDU 
is still active resulting in mild F enhancements at the stellar surface. 
This explains why while the most prolific AGB stars in terms of fluorine 
production have initial masses around 2 \msun, F production still occurs 
at slightly higher masses. On the other hand, more massive AGB models 
($>$5 \msun)~experience hotter HBB and thus more efficient F 
destruction as well as higher temperatures in the thermal pulses activating 
also $^{19}$F($\alpha$,p)$^{22}$Ne reactions. This, combined with  
fewer final TDU episodes when HBB has ceased means that they do not
replenish F at the stellar surface. 

Interestingly, the same conclusion is drawn by \cite{roederer11} by 
exploring the heavy-element ratios, [$hs$/$ls$] and [Pb/$hs$]. 
Comparing their abundances with models by \cite{roederer10}, these 
authors deduced that the low mass AGBs ($\leq$3\msun) cannot account for 
the observed trend. More importantly, they concluded that a match to the 
$s$-process element abundances is provided by the 5 \msun\ AGB model, and 
this is confirmed by checking the models of \cite{lugaro12}. 
Furthermore, and very interestingly, these models predict Na and C production in agreement
with the observations.

{\em Our result provides a further, independent confirmation to this 
previous hint: indications from both light (fluorine here) and heavy 
elements converge toward the AGB stars of the same mass 
as the best candidate polluters, indicating 
that the age difference between the two sub-groups in M22 cannot be 
larger than a few hundreds Myr.} It should be mentioned
that by analyzing the double sub-giant branch (SGB) of this 
cluster, \cite{marino12} concluded that the age spread can be {\em at 
most} $\sim$300 Myr.

The fact that three independent studies, involving different and 
complementary techniques/approaches, produce the same result is 
encouraging. However, a comprehensive understanding of the whole picture 
is still missing. As also stressed by \cite{roederer11}, 
if relatively massive AGBs (4-5 \msun)~produced the $s$-process 
elements, and if these stars are also responsible for the observed 
$p$-capture element anticorrelations, then it is not explained that 
an $s$-process enrichment is present in M22 but is not associated 
with O and Na abundance anomalies, nor it is seen in any other GCs 
where the Na-O anticorrelations are observed. 

On the other hand, it is clear from the observed F 
abundance trends that we must select slightly more massive AGBs, i.e., $\geq$ 6\msun\, if we wish to explain the 
light element variations in GCs, since 
this is the AGB mass range where F and O can be destroyed by HBB 
resulting in the observed F-O correlation. We might tentatively suggest 
that perhaps the production of the $s$-process elements in AGB models 
with initial mass $\geq$ 6\msun~is less efficient than currently predicted. 
This could be the result of a stronger mass-loss rate or a less 
efficient TDU in this mass range than those employed in the models by 
\cite{lugaro12}. This possibility is well within model uncertainties 
and need to be investigated; in a forthcoming paper 
(D'Orazi et al., in preparation) we will attack these issues, presenting new observations and AGB models 
and discussing their strength/weakness in reproducing the observed abundance trends in GCs.

Alternatively, we may conclude that 
massive AGBs are not the inter-cluster polluters; however 
several lines of evidence point to those stars, such as the need for a Li 
production between first and second generation stars (as in the case of 
M4, \citealt{dm10}; \citealt{mucciarelli11}; \citealt{monaco12}). Further efforts, from both observational and 
theoretical perspectives, are needed; in particular the lack of a 
complete set of models for AGB stars and massive rotating stars 
with different mass and metallicity, following the whole nucleosynthetic path from Li to Pb, still hampers a 
quantitatively robust comparison between theory and observations.

\section{Summary and concluding remarks}\label{sec:summary}

We presented fluorine abundances for a sample of six RGB 
stars belonging to the metal-poor globular cluster M22. The sample was 
selected to include $s$-process-rich and $s$-process-poor 
stars, as defined in MSK11. In addition, within each of these 
cluster sub-components, we targeted both O-rich (Na-poor) and O-poor 
(Na-rich) stars. 

We gathered evidence of the presence of a F-O 
correlation and of an F-Na anticorrelation.
Such chemical pattern, notably revealed in each cluster 
sub-group, is in agreement with F destruction during the hot H burning: 
fluorine follows the same trend defined by O, Na, C, N, Mg, Al, as predicted by the 
multiple population scenario. 

Most interestingly, we found that 
the $s$-process- (metal-) rich component is also characterized by a 
larger F content than the $s$-process (metal-) poor component. The 
comparison between our observations and AGB models points to stars with 
masses around 4-5 \msun~as responsible for such a trend, corroborating 
previous hints by \cite{roederer11} and \cite{marino12}, and confirming 
that the age spread across the two different stellar generations in M22 
cannot be larger than a few hundreds Myr.

\acknowledgments
The publication made extensive use of the NASA ADS and SIMBAD databases.
We kindly acknowledge B. Plez for having provided his unpublished line lists.
This work was partially funded by PRIN INAF 2009 (``Formation and Early
Evolution of Massive Star Clusers'') and PRIN INAF 2011 (``Multiple populations in globular clusters: their role in the Galaxy assembly'').
V.D. thanks Kais Hamza for very useful discussions.

%% In a manner similar to \objectname authors can provide links to dataset
%% hosted at participating data centers via the \dataset{} command.  The
%% second curly bracket argument is printed in the text while the first
%% parentheses argument serves as the valid data set identifier.  Large
%% lists of data set are best provided in a table (see Table 3 for an example).
%% Valid data set identifiers should be obtained from the data center that
%% is currently hosting the data.
%%
%% Note that AASTeX interprets everything between the curly braces in the 
%% macro as regular text, so any special characters, e.g. "#" or "_," must be 
%% preceded by a backslash. Otherwise, you will get a LaTeX error when you 
%% compile your manuscript.  Special characters do not 
%% need to be escaped in the optional, square-bracket argument.

\clearpage

%% the file, uncomment the \plotone, \plottwo, and
%% If you need a layout that cannot be achieved with \plotone or
%% \plottwo, you can invoke the graphicx package directly with the
%% \includegraphics command or use \plotfiddle. For more information,
%% please see the tutorial on "Using Electronic Art with AASTeX" in the
%% documentation section at the AASTeX Web site,
%% http://www.journals.uchicago.edu/AAS/AASTeX.
%%
%% The examples below also include sample markup for submission of
%% supplemental electronic materials. As always, be sure to check
%% the instructions to authors for the journal you are submitting to
%% for specific submissions guidelines as they vary from

%% This example uses \plotone to include an EPS file scaled to
%% 80% of its natural size with \epsscale. Its caption
%% has been written to indicate that additional figure parts will be
%% available in the electronic journal.

\clearpage


\begin{thebibliography}{}
\bibitem[\protect\citeauthoryear{Abia et al.}{2009}]{abia09} Abia, C., Recio-Blanco, A., de Laverny, P., et al. 2009, ApJ, 694, 971
\bibitem[\protect\citeauthoryear{Abia et al.}{2010}]{abia10} Abia, C., Cunha, K., Cristallo, S., et al. 20110, ApJ, 715, L94
\bibitem[\protect\citeauthoryear{Abia et al.}{2011}]{abia11}Abia, C., Cunha, K., Cristallo, S., et al. 2011, ApJ, 737, 8 
\bibitem[\protect\citeauthoryear{Alves-Brito et al.}{2012}]{abrito12} Alves-Brito, A., Melendez, J., Vasquez, S., Karakas, A.I. 2012, A\&A, 540, A3 
\bibitem[\protect\citeauthoryear{Anders \& Grevesse}{1989}]{ag89} Anders, E., Grevesse, N. 1989, Geochim. Cosmochim. Acta, 53, 197 
\bibitem[\protect\citeauthoryear{Asplund et al.}{2005}]{asplund05} Asplund, M., Grevesse, N., Sauval, A.J. 2005, 
in Cosmic Abundances as Records of Stellar Evolution and Nucleosynthesis, ed. T.G. Barnes, III \& F.N. Bash, PASP, 25

\bibitem[\protect\citeauthoryear{Brown et al.}{1990}]{brown90} Brown, J.A., Wallerstein, G., Oke, J.B. 1990, AJ, 100, 1561
\bibitem[\protect\citeauthoryear{Brown \& Wallerstein}{1992}]{bw92} Brown, J.A., Wallerstein, G. 1992, AJ, 104, 1818 
\bibitem[\protect\citeauthoryear{Burris et al.} {2000}]{burris00} Burris, D.L., Pilachowski, C.A., Armandroff, T.E., et al. 2000, ApJ, 544, 302
\bibitem[\protect\citeauthoryear{Busso et al.}{1999}]{busso99} Busso, M., Gallino, R., Wasserburg, G.J. 1999, ARA\&A, 37, 239 
\bibitem[\protect\citeauthoryear{Carretta et al.} {2009a}]{carretta09a} Carretta, E., Bragaglia, A., Gratton, R.G., et al. 2009a, A\&A, 505, 117
\bibitem[\protect\citeauthoryear{Carretta et al.} {2009b}]{carretta09b} Carretta,  E., Bragaglia, A., Gratton, R.G., Lucatello, S. 2009b, A\&A, 505, 139
\bibitem[\protect\citeauthoryear{Carretta et al.} {2009c}]{carretta09c}	Carretta, E., Bragaglia, A., Gratton, R.G., D'Orazi, V., Lucatello, S. 2009c, A\&A, 508, 695
\bibitem[\protect\citeauthoryear{Carretta et al.} {2010a}]{carretta10a} Carretta, E., Gratton, R.G., Lucatello, S., et al. 2010a, ApJ, 722, L1
\bibitem[\protect\citeauthoryear{Carretta et al.} {2010b}]{carretta10b} Carretta, E., Bragaglia, A., Gratton, R.G., et al. 2010b, A\&A, 520, 95
\bibitem[\protect\citeauthoryear{Chiappini et al.} {2011}]{chiappini11} Chiappini, C., Frischknecht, U., Meynet, G., et al. 2011, Nature, 474, 666
\bibitem[\protect\citeauthoryear{Cohen et al.}{2011}]{cohen11} 	Cohen, J.G., Huang, W., Kirby, E.N. 2011, ApJ, 740, 60
\bibitem[\protect\citeauthoryear{Cunha et al.}{2003}]{cunha03} Cunha, K., Smith, V.V., Lambert, D.L., Hinkle, K.H. 2003, AJ, 1026, 305 
\bibitem[\protect\citeauthoryear{Da Costa et al.}{2009}]{dacosta09} Da Costa, G.S., Held, E.V., Saviane, I., Gullieuszik, M. 2009, ApJ, 705, 1481 
\bibitem[\protect\citeauthoryear{Da Costa \& Marino}{2010}]{dacosta10} Da Costa, G.S., Marino A.F. 2010, PASA, 28, 28 
\bibitem[\protect\citeauthoryear{D'Antona et al.}{1983}]{dantona83} D'Antona, F., Gratton, R., Chieffi, A. 1983, Memorie della Societ\'a
Astronomica Italiana, 54, 173D
\bibitem[\protect\citeauthoryear{Decin}{2000}]{decin00} Decin, L. 2000, Ph.D. Thesis, Catholique University of Leuven Department of Physics and Astronomy
\bibitem[\protect\citeauthoryear{Decressin et al.}{2007}]{decressin07} Decressin, T., Meynet, G., Charbonnel, C., Prantzos, N., Ekstr\"{o}m, S. 2007, A\&A, 464, 1029
\bibitem[\protect\citeauthoryear{de Mink et al.}{2009}]{demink09} de Mink, S.E., Pols, O.R., Langer, N., Izzard, R.G. 2009, A\&A, 507, L1
\bibitem[\protect\citeauthoryear{D'Orazi et al.}{2010}]{dorazi10} D'Orazi, V., Gratton, R.G.., Lucatello, S., et al. 2010, ApJ, 719, L213
\bibitem[\protect\citeauthoryear{D'Orazi \& Marino}{2010}]{dm10} D'Orazi, V., Marino, A.F. 2010, ApJ, 716, L166
\bibitem[\protect\citeauthoryear{Ferraro et al.}{2009}]{ferraro09} Ferraro, F.R., Dalessandro, E., Mucciarelli, A., et al. 2009, Nature, 462, 483
\bibitem[\protect\citeauthoryear{Forestini et al.}{1992}]{forestini92} Forestini, M., Goriely, S., Jorissen, A., Arnould, M. 1992, A\&A, 261, 157
\bibitem[\protect\citeauthoryear{Gratton et al.}{2012}] {gratton12} Gratton, R.G., Carretta, E., Bragaglia, A. 2012, A\&ARv, 20, 50
\bibitem[\protect\citeauthoryear{Gustafsson et al.}{2008}]{gustafsson08} Gustaffson, B., Edvardsson, B., Eriksson, K., et al. 2008, A\&A, 486, 951 
\bibitem[\protect\citeauthoryear{Harris}{1996}]{harris96} Harris, W.E. 1996, AJ, 112, 1487
\bibitem[\protect\citeauthoryear{Hinkle et al.}{1995}]{hinkle} Hinkle, K., Wallace, L., Livingston,W. C. 1995,
“Infrared atlas of the Arcturus spectrum, 0.9-5.3 microns”, eds. K. Hinkle, L. Wallace, and W. C. Livingston (San Francisco: ASP) ISBN: 1-866733-04-X
\bibitem[\protect\citeauthoryear{James et al.}{2004}]{james04} James, G., Fran\c{c}ois, P., Bonifacio, P., et al. 2004, A\&A, 427, 825
\bibitem[\protect\citeauthoryear{Johnson \& Pilachowski}{2010}]{jp10} Johnson, C.I., Pilachowski, C.A. 2010, ApJ, 722, 1373 
\bibitem[\protect\citeauthoryear{Jorissen et al.}{1992}]{jorissen92} Jorissen, A., Smith, V.V., Lambert, D. 1992, A\&A, 261, 164
\bibitem[\protect\citeauthoryear{Kaeufl et al.}{2004}]{kaeufl04} Kaeufl, H.-U., et al. 2004, Proc. SPIE, 5492, 1218
\bibitem[\protect\citeauthoryear{Karakas \& Lattanzio}{2003}]{karakas03} Karakas, A.I., Lattanzio, J.C. 2003, PASA, 20, 279
\bibitem[\protect\citeauthoryear{Kayser et al.}{2008}]{kayser08} Kayser, A., Hilker, M., Grebel, E.K., Willemsen, P.G. 2008, A\&A, 486, 437 
\bibitem[\protect\citeauthoryear{Kurucz}{1993}]{kur93} Kurucz, R. 1993, CD-ROM 13, ATLAS9 Stellar Atmosphere Programs and 2 km/s Grid (Cambridge: SAO), 13
\bibitem[\protect\citeauthoryear{Lind et al.}{2011}]{lind11} Lind, K., Asplund, M., Barklem, P.S., Belyaev, A.K. 2011, A\&A, 528, A103
\bibitem[\protect\citeauthoryear{Lucatello et al.}{2011}]{lucatello11} Lucatello, S., Masseron, T., Johnson, J.A., Pignatari, M., Herwing, F. 2011, ApJ, 792, 40
\bibitem[\protect\citeauthoryear{Lugaro et al.}{2004}]{lugaro04} Lugaro, M., Ugalde, C., Karakas, A.I., et al. 2004, ApJ, 615, 934
\bibitem[\protect\citeauthoryear{Lugaro et al.}{2012}]{lugaro12} Lugaro, M., Karakas, A.I., Stancliffe, R.J., Rijs, C. 2012, ApJ, 747, 2
\bibitem[\protect\citeauthoryear{Maccarone \& Zurek}{2012}]{maccarone12} Maccarone, T.J., Zurek, D.R. 2012, MNRAS, 423, 2 
\bibitem[\protect\citeauthoryear{Marino et al.}{2009}]{marino09} Marino, A. F., Milone, A. P., Piotto, G., et al. 2009, A\&A, 505, 1099
\bibitem[\protect\citeauthoryear{Marino et al.}{2011a}] {marino11a}Marino, A. F., Milone, A. P., Piotto, G., et al. 2011a, ApJ, 731, 64 
\bibitem[\protect\citeauthoryear{Marino et al.}{2011b}] {marino11b}Marino, A. F., Sneden, C., Kraft, R.P., et al. 2011b, A\&A, 532, A8 (MSK11)
\bibitem[\protect\citeauthoryear{Marino et al.}{2012}]{marino12} Marino, A.F., Milone, A., Sneden, C., et al. 2012, A\&A, 541, 15
\bibitem[\protect\citeauthoryear{Meynet \& Arnould}{2000}]{meynet00} Meynet, G., Arnould, M. 2000, A\&A, 355, 176 
\bibitem[\protect\citeauthoryear{Mowlavi et al.}{1998}]{mowlavi98} Mowlavi, N., Jorissen, A., Arnould, M. 1998, A\&A, 334, 153
\bibitem[\protect\citeauthoryear{Monaco et al.}{2012}]{monaco12} Monaco, L., Villanova, S., Bonifacio, P., et al. 2012, A\&A, 539, 157
\bibitem[\protect\citeauthoryear{Mucciarelli et al.}{2011}]{mucciarelli11} Mucciarelli, A., Salaris, M., Lovisi, L., et al. 2011, MNRAS, 412, 81
\bibitem[\protect\citeauthoryear{Mucciarelli et al.}{2012}]{mucciarelli12} Mucciarelli, A., Bellazzini, M., Ibata, R., et al. 2012, MNRAS, 426, 2889 
\bibitem[\protect\citeauthoryear{Norris \& Freeman}{1983}]{norris83} Norris, J., Freeman, K.C. 1983, ApJ, 266, 130
\bibitem[\protect\citeauthoryear{Palacios et al.}{2005}]{palacios05} Palacios, A., Arnould, M., Meynet, G. 2005, A\&A, 443, 243
\bibitem[\protect\citeauthoryear{Pignatari et al.}{2008}]{pignatari08} 	Pignatari, M., Gallino, R., Meynet, G., et al. 2008, ApJ, 687, L95
\bibitem[\protect\citeauthoryear{Pignatari et al.}{2010}]{pignatari10} Pignatari, M., Gallino, R., Heil, M., et al. 2010, ApJ, 710, 1557
\bibitem[\protect\citeauthoryear{Pilachowski et al.}{1982}]{pilachowski82} Pilachowski, C.A., Leep, E.M., Wallerstein, G.J., Peterson, R.C. 1982, ApJ, 263, 187 
\bibitem[\protect\citeauthoryear{Raiteri et al.}{1993}]{raiteri93} Raiteri, C.M., Gallino, R., Busso, M., Neuberger, D., Kaeppeler, F. 1993, ApJ, 419, 207
\bibitem[\protect\citeauthoryear{Ram\'{i}rez \& Allende Prieto}{2011}]{ramirez11} Ram\'{i}rez, I., Allende Prieto, C. 2011, ApJ, 743, 135
\bibitem[\protect\citeauthoryear{Roederer et al.}{2010}]{roederer10} Roederer, I.U., Cowan, J.J., Karakas, A.I., et al. 2010, 
724, 925
\bibitem[\protect\citeauthoryear{Roederer et al.}{2011}]{roederer11} Roederer, I.U., Marino, A.F., Sneden, C. 2011, ApJ, 742, 37
\bibitem[\protect\citeauthoryear{Ryde et al.}{2010}]{ryde10} Ryde, N., Gustafsson, B., Edvardsson, B., et al. 2010, A\&A, 509, 20
\bibitem[\protect\citeauthoryear{Smith \& Kraft}{1996}]{smith96} Smith, G.H., Kraft, R.P 1996, PASP, 108, 344
\bibitem[\protect\citeauthoryear{Smith et al.}{2005}]{smith05} Smith, V.V., Cunha, K., Ivans, I.I., et al. 2005, ApJ, 633, 392 
\bibitem[\protect\citeauthoryear{Smith}{2008}]{smith08} Smith, G.H. 2008, PASP, 120, 952
\bibitem[\protect\citeauthoryear{Sneden}{1973}]{sneden73} Sneden, C. 1973, ApJ, 184, 839
\bibitem[\protect\citeauthoryear{Ventura et al.}{2001}]{ventura01} Ventura, P., D'Antona, F., Mazzitelli, I., Gratton, R.G. 2001, ApJ, 550, L65
\bibitem[\protect\citeauthoryear{Yong \& Grundahl}{2008}]{yong08} Yong, D., \& Grundahl, F. 2008, ApJ, 672, L29
\bibitem[\protect\citeauthoryear{Yong et al.}{2008}]{yong+08} Yong, D., Melendez, J., Cunha, K., et al. 2008, ApJ, 689, 1020
\bibitem[\protect\citeauthoryear{Woosley et al.}{1990}]{woosley90}Woosley, S.E., Hartmann, D.H., Hoffman, R.D., Haxton, W.C. 1990, ApJ, 356, 272 

\end{thebibliography}
\end{document}